\documentclass[11pt]{article}
\usepackage{graphicx}
\usepackage{ifthen}

\newboolean{physMode}
\setboolean{physMode} {true}

\newboolean{jourMode}
\setboolean{jourMode} {true}

\newboolean{confFlag}
\setboolean{confFlag} {true}

\newboolean{authFlag}
\setboolean{authFlag} {true}

\newcommand{\BABARPubYear}    {06}

\newcommand{\BABARConfNumber}{042}
\newcommand{\SLACPubNumber} {12033}

\input{pubboard/babarsym}
 
\long\def\inst#1{\par\nobreak\kern 4pt\nobreak
    {\it #1}\par\vskip 10pt plus 3pt minus 3pt}

\newcommand{\pid}     {PID}

\newcommand{\belle} {Belle}

\newcommand{\Cherenkov} {Cherenkov}

\newcommand{\geantfour} {GEANT4}

\newcommand{\ROOT} {ROOT}

\newcommand{\deltaE} {\mbox{$\Delta E$}}

\newcommand{\pdf} {\mbox{\rm PDF}}

\newcommand{\like} {\ensuremath{\cal{L}}}

\newcommand{\rms}  {\mbox{\rm RMS}}

\newcommand{\mlp}   {\ensuremath{m(\Lambda p)}}

\newcommand{\mlbp}   {\ensuremath{m(\bar{\Lambda} p)}}

\newcommand{\splot} {\ensuremath{\mbox{}_s\cal{P}}lot}

\newcommand{\TotalOnPeakLumi}      {210.3}

\newcommand{\TotalBPairs}          {232}

\newcommand{\TotalBPairsPerc}      {$1.1\%$}

\newcommand{\DeltaEWinCut}     {27}

\newcommand{\MesLowWinCut}     {5.274}

\newcommand{\PiLHLoosePiKLikeRatioMin} {0.22}
\newcommand{\PiLHLoosePiPLikeRatioMin} {0.02}

\newcommand{\LambdaInvMassResOne}      {0.7}
\newcommand{\LambdaInvMassResTwo}      {1.3}
\newcommand{\LambdaInvMassResThree}    {4.0}
\newcommand{\LambdaInvMassResMean}     {1.8}

\newcommand{\GoodTracksLoosePtMin}     {100}

\newcommand{\FishermcsigEffNoErr}         {$71.2\%$}
\newcommand{\FisherbkgEffNoErr}           {$6.4\%$}

\newcommand{\VtxProbCutL}                    {\mbox{$10^{-6}$}}

\newcommand{\LambdaInvMassCutL}          {1.111}
\newcommand{\LambdaInvMassCutU}          {1.121}

\newcommand{\LambdaFlightSigCutL}        {35}

\newcommand{\LmFlCutdataSideRejNoErr}   {$42\%$}
\newcommand{\LmFlCutmcsigSideEffNoErr}   {$90\%$}

\newcommand{\DeltaEMesSigCorrelation}            {-7.4\%}

\newcommand{\DeltaEMesBkgCorrelation}             {-0.5\%}

\newcommand{\SCFExpectedFractionNoErr}      {$0.59\%$}

\newcommand{\LambdaCInvMassCut}          {20}
\newcommand{\LambdaCInvMassCutSigma}     {5}

\newcommand{\DeltaEFitRegion}               {200}
\newcommand{\MesFitRegion}                  {5.2}
\newcommand{\MesFitRegionLow}               {5.2}
\newcommand{\MesFitRegionUp}                {5.29}

\newcommand{\FDCutSystematicPerc}    {$2.4\%$}

\newcommand{\VtxCutSystematicPerc}      {$5.0\%$}

\newcommand{\BZeroOverBChargedFraction}  {$49.3\pm 0.8\%$}
\newcommand{\BZeroOverBChargedDefault}   {$50\%$}

\newcommand{\TotalBPairsSystematic}      {\TotalBPairsPerc}

\newcommand{\TotalTrackSystematic}       {3.9\%}
\newcommand{\PIDSystematic}              {1.4\%}
\newcommand{\DalitzStatSystematic}       {2.0\%}
\newcommand{\LambdaFlightSystematic}     {2.8\%}

\newcommand{\LambdaMassSystematic}       {2.4\%}
\newcommand{\SignalPDFSystematic}        {3.2\%}
\newcommand{\BackgroundPDFSystematic}    {2.2\%}
\newcommand{\LikelihoodParameters}       {3.9\%}
\newcommand{\DeltaESystematic}           {1.7\%}
\newcommand{\SCFSystematic}              {0.8\%}
\newcommand{\LikelihoodBias}             {0.6\%}
\newcommand{\LambdaCPiResBackSystematic} {0.2\%}
\newcommand{\BLambdaPPiVetoSystematic}   {0.5\%}
\newcommand{\LambdaCVetoSystematic}      {\BLambdaPPiVetoSystematic}
\newcommand{\LambdaPPiBFSystematic}      {0.8\%}
\newcommand{\BFractionSystematic}        {1.4\%}
\newcommand{\MesSystematic}              {1.9\%}
\newcommand{\DESystematic}               {1.3\%}
\newcommand{\TotalSystematic}            {9.4\%}

\newcommand{\DeltaEScaleDiffPerc}        {5\%}

\newcommand{\DataSampleSelectedEvents}      {4261}

\newcommand{\LikeNSFitResult}          {$73.7^{+12.0}_{-11.2}$}
\newcommand{\LikeNBFitResult}          {$4187\pm66$}
\newcommand{\LikeMuDEFitResult}        {$-1.71\pm 3.10$} %
\newcommand{\LikeCOneDEFitResult}      {$-3.71\pm 0.25$}
\newcommand{\LikeMuMesFitResult}       {$5.2808\pm 0.0004$} %
\newcommand{\LikeCArgusMesFitResult}   {$-15.1\pm 1.7$}

\newcommand{\SPlotEffCorrectedYield}   {$488\pm 79$}

\newcommand{\FinalBFVal}           {3.30}   %
\newcommand{\FinalBFErrStat}       {0.53}   %
\newcommand{\FinalBFErrSyst}       {0.31}   %
\newcommand{\FinalBFCombined}      {\mbox{$\left[
\FinalBFVal\pm\FinalBFErrStat\ ({\rm stat.})\pm\FinalBFErrSyst\ ({\rm syst.})\right]
\times 10^{-6}$}}

\begin{document}
{\pagestyle{empty}

\begin{flushright}
\babar-CONF-\BABARPubYear/\BABARConfNumber \\
SLAC-PUB-\SLACPubNumber \\
\end{flushright}

\par\vskip 5cm

\begin{center}
\Large \bf Measurement of the $B^0\rightarrow \bar{\Lambda} p \pi^-$ Branching
Fraction and Study of the Decay Dynamics
\end{center}
\bigskip

\begin{center}
\large The \babar\ Collaboration\\
\mbox{ }\\
\today
\end{center}
\bigskip \bigskip

\begin{center}
\large \bf Abstract
\end{center}
We present a  measurement of the $B^0\rightarrow \bar{\Lambda} p \pi^-$
branching fraction performed using the \babar\ detector at the \pep2\
asymmetric energy \epem\ collider.
Based on a \TotalBPairs\ million $B\bar{B}$ pairs data sample we measure:
$
{\cal{B}}(B^0\rightarrow \bar{\Lambda} p \pi^-) = \FinalBFCombined
.$
 A measurement of the differential spectrum as a function of the di-baryon
invariant mass \mlp\ is also presented; this shows a near-threshold
enhancement similar to that observed in other baryonic B decays.
\vfill
\begin{center}

Submitted to the 33$^{\rm rd}$ International Conference on High-Energy Physics, ICHEP 06,\\
26 July---2 August 2006, Moscow, Russia.

\end{center}

\vspace{1.0cm}
\begin{center}
{\em Stanford Linear Accelerator Center, Stanford University, 
Stanford, CA 94309} \\ \vspace{0.1cm}\hrule\vspace{0.1cm}
Work supported in part by Department of Energy contract DE-AC03-76SF00515.
\end{center}

\newpage
} %

\ifthenelse{\boolean{authFlag}}{
\begin{center}
\small

The \babar\ Collaboration,
\bigskip

%
{B.~Aubert,}
{R.~Barate,}
{M.~Bona,}
{D.~Boutigny,}
{F.~Couderc,}
{Y.~Karyotakis,}
{J.~P.~Lees,}
{V.~Poireau,}
{V.~Tisserand,}
{A.~Zghiche}
\inst{Laboratoire de Physique des Particules, IN2P3/CNRS et Universit\'e de Savoie,
 F-74941 Annecy-Le-Vieux, France }
% ...

\end{center}\newpage

}{}

\ifthenelse{\boolean{confFlag}}{
\section{INTRODUCTION}
}{
\section{Introduction}
}

\label{sec:Introduction}

Observations of charmless three-body baryonic B decays have been reported
recently by both the \babar\ and \belle\ collaborations
\cite{ref:Belle_Blm_Observ,ref:Belle_Blm_Update,ref:BaBar_Bppk_Study}. A
common feature of these decay modes is the peaking of the
baryon-antibaryon mass spectrum toward threshold. This feature has
stimulated considerable interest among theorists as a key element in the
explanation of the unexpectedly high branching fractions for these decays
\cite{ref:Hou_Soni_PRL86,ref:Chua_Hou_EurC29}.
We report a measurement of the branching fraction for $B^0$ decay to
the $\bar{\Lambda} p \pi^-$ final state \footnote{Inclusion of the charge 
conjugate mode is implied.}.
 In the Standard Model this decay proceeds through
tree level $b\rightarrow u$ and penguin $b\rightarrow s$ amplitudes. It is of
interest to study the structure of the decay amplitude in the Dalitz plane
and to test the afore-mentioned theoretical expectations.
This channel may also be used to search for direct CP violation, and
with the 
$\Lambda$ hyperon in the final state, its spin self-analyzing weak
decay to $p\ \pi$, may be used, with increased statistics, to study
the chirality structure of weak
$b\rightarrow s$ transitions \cite{ref:Suzuki_JPG29} and to
 probe T violation \cite{ref:Hou_Soni_PRL86,ref:Geng_Hsiao_JMPA}.

\ifthenelse{\boolean{jourMode}}{}{
 T-naive is distinct from ordinary T symmetry, in that it doesn't exchange
initial and final states, while still reversing the sign of momenta and
angular momenta. Despite this difference, studying the former with triple
T-odd product asymmetries, that are measurable in the considered baryonic
B decay, could provide insight into the latter~\cite{ref:Geng_Hsiao_JMPA}.
}

\ifthenelse{\boolean{confFlag}}{
\section{THE \babar\ DETECTOR AND DATASET}
}{
\section{Dataset and Selection}
}

\label{sec:Dataset}

The data sample consists of \TotalBPairs\ million $B\bar{B}$ pairs
corresponding to an integrated luminosity of \TotalOnPeakLumi\ ${\rm
fb^{-1}}$, collected at the \FourS\ resonance with the \babar\ detector.
The detector is described in detail elsewhere~\cite{ref:BaBar_NIM}.  
Charged particle trajectories are measured in a tracking system consisting
of a five-layer double-sided silicon vertex tracker (SVT) and a 40-layer
central drift chamber (DCH), both operating in a 1.5-T axial magnetic
field.  A ring-imaging Cherenkov detector (DIRC)  
is used for charged-particle identification.
A CsI(Tl) electromagnetic calorimeter (EMC) is used to
detect and identify photons and electrons, while muons are identified in
the instrumented flux return of the magnet (IFR). A \babar\ detector Monte
Carlo simulation based on \geantfour~\cite{ref:geantfour} is used to optimize selection
criteria and determine selection efficiencies.

\ifthenelse{\boolean{confFlag}}{
\section{EVENT SELECTION}
\label{sec:EventSelection}
}{
}

We reconstruct $\Lambda$ candidates in the $\Lambda\rightarrow p\pi$ decay
mode as combinations of oppositely-charged tracks, assigned the proton
and pion mass hypotheses, and fit to a common vertex. A fit to the invariant
mass distribution of reconstructed candidates with a triple Gaussian
 gives
\rms\ widths of \LambdaInvMassResOne\mevcc, \LambdaInvMassResTwo\mevcc and
\LambdaInvMassResThree\mevcc for the narrow, intermediate and wide Gaussians
respectively, with an average value of \LambdaInvMassResMean\mevcc.
Combinations with an invariant mass in the range \LambdaInvMassCutL\ --
\LambdaInvMassCutU\gevcc\ are refit with a mass constraint to the
nominal $\Lambda$ mass \cite{ref:PDG}, and combined with two additional tracks with
opposite charges, each with momentum transverse to the beam greater than
\GoodTracksLoosePtMin\mevc.

Measurements of the average energy loss
(\dedx) in the tracking devices, angle of the \Cherenkov\ cone in the
DIRC, and energy releases in the EMC and IFR are combined to give a
likelihood estimator for a track to be consistent with a given particle
hypothesis. Tracks with likelihood ratios satisfying the very loose
particle identification (\pid) criterion
$L_p/L_K>1/3$ and $L_p/L_\pi>1$ are assumed to be protons. In
addition, the pion that originates from the B decay vertex must
satisfy a loose \pid\ criterion $L_\pi/L_K>\PiLHLoosePiKLikeRatioMin$ and
$L_\pi/L_p>\PiLHLoosePiPLikeRatioMin$. A Kalman  fit~\cite{ref:Hulsbergen_Kalman} to the full decay
sequence is used to reconstruct the B vertex; only candidates with a fit
probability $P_{\rm vtx}>\VtxProbCutL$ are considered.

The primary background to the reconstructed decay channel arises
from light quark
continuum events $e^+e^-\rightarrow q\bar{q}$ ( $q = u,d,s,c$ ),
which are characterized by collimation of
final state particles with respect to the quark direction, in contrast
to the more spherical $B\bar{B}$ events.
Exploiting these different topologies we can increase the signal significance
using topological variables computed from the center-of-mass (CM) momenta
of all reconstructed charged and neutral particles in the event. For each
event we linearly combine the sphericity, the angle between the B thrust
axis and detector longitudinal axis, and the zeroth and second order
Legendre moments into a Fisher discriminant ($\cal F$) \cite{ref:Fisher},
whose coefficients are chosen to optimize the separation of signal and
continuum background Monte Carlo samples. After optimization of the
selection with respect to the simultaneous variation of all the selection
criteria, we obtain that the Fisher requirement retains
\FishermcsigEffNoErr\ of the candidates from the signal Monte Carlo sample
and only \FisherbkgEffNoErr\ from the continuum background Monte Carlo.

To further reduce combinatoric background we take advantage of the long mean
lifetime of $\Lambda$ particles and require that the separation of the
$\Lambda$ and $B^0$ vertices divided by its measurement error, computed on
a per candidate basis by the fit procedure, exceeds \LambdaFlightSigCutL.
This criterion was optimized on Monte Carlo events and is effective
in rejecting \LmFlCutdataSideRejNoErr\ of
combinatorial background that survived all other cuts, 
while retaining \LmFlCutmcsigSideEffNoErr\ of signal candidates.
The only sizable B background is from the process
$B^0\rightarrow
\bar{\Lambda_c}^-\left(\rightarrow\bar{\Lambda}\pi^-\right)p$,
and we reject this
with a veto on candidates with an invariant mass $m(\Lambda\pi)$,
within \LambdaCInvMassCut\mevcc,
approximately \LambdaCInvMassCutSigma\ standard deviations, of the
nominal $\Lambda_c$ mass \cite{ref:PDG}.

The kinematic constraints on B mesons produced at the \FourS allow
further background discrimination using the variables \mes\ and \deltaE.
We define
 $
\mes = \sqrt{\left(\frac{s}{2}+\vec{p}_i\cdot\vec{p}_B\right)^2/E_i^2 -
	\left.\vec{p}_B\right.^{2}}
$
 where $(E_i,\vec{p}_i)$ is the four momentum of the initial $e^+e^-$
system and $\vec{p}_B$ the momentum of the reconstructed B candidate,
both measured in the laboratory frame, and $s$ is the square of the
total available energy in the \FourS\ center of mass frame. We have
 $
\deltaE = E^*_B - \frac{\sqrt s}{2}
$
 where $E^*_B$ is the B energy in the \FourS\ center of mass frame.
Candidates satisfying $|\deltaE|<\DeltaEFitRegion\mev$ and
$\MesFitRegionLow<\mes<\MesFitRegionUp\gevcc$ are used in the maximum
likelihood fitting process.

\ifthenelse{\boolean{confFlag}}{
\section{BRANCHING FRACTION}
}{
\section{Branching Fraction}
}

\label{ssec:sPlot_fit}

We perform the measurement using a maximum-likelihood fit on \mes-\deltaE\ 
observables of reconstructed B candidates. The \splot\ 
technique\ifthenelse{\boolean{physMode}}{ \cite{ref:Pivk_sPlot} }{
\cite{ref:Pivk_sPlot,ref:Pivk_sPlot_article,ref:Cahn_sPlot,ref:Snyder_sPlot}
}, is then used to determine the \mlbp\ distribution of
reconstructed candidates and, once the correction for the nonuniform
reconstruction efficiency is applied, measure the \mlbp
-dependent differential rate together with the total branching fraction.

We consider as signal events only reconstructed B meson candidates whose
daughters are correctly assigned in the decay chain. By self-cross-feed, we
refer to candidates reconstructed as signal events in which one or more
particles are not correctly assigned in the decay chain. Examples of such
misreconstruction include events in which a proton from the other B meson
are associated to the signal B, and  events where the protons
from the signal B and $\Lambda$ decays are interchanged.
 We define the total \pdf\ in the \deltaE-\mes\ plane as the sum
of signal, self-cross-feed, and background components:
\ifthenelse{\boolean{confFlag}}{
\begin{eqnarray}
}{
\begin{eqnarray*}
}
{\like}=\frac{1}{N!}e^{-(N_{\rm S} + N_{\rm B} + N_{\rm S} f_{\rm scf})}
\prod_{\alpha=1}^{N} \left[N_{\rm S}{\cal{P}}_{\rm S,\alpha}
+ N_{\rm B}{\cal{P}}_{\rm B,\alpha}
\ifthenelse{\boolean{confFlag}}{}{
\right.\\\left.}
 + N_{\rm S}f_{\rm scf}{\cal{P}}_{\rm scf,\alpha}\right]
\ifthenelse{\boolean{confFlag}}{
\end{eqnarray}
}{
\end{eqnarray*}
}
 where the product is over the $N$ fitted events with $N_{\rm S}$ and
$N_{\rm B}$ representing the number of signal and background candidates
and $f_{\rm scf}$ representing the self-cross-feed fraction. The three
$\cal{P}$ functions are taken as products of 1-dimensional \deltaE\ and
\mes\ \pdf's.
 We are justified in this simplification by the small
correlation between these two variables in our Monte Carlo sample,
measured as \DeltaEMesSigCorrelation\ for signal, and
\DeltaEMesBkgCorrelation\ for background. The \mes\ \pdf\ is taken as a
double Gaussian for the signal and a threshold function
\cite{ref:Argus} for the background. The \deltaE\ \pdf\ is taken as a
double Gaussian for the signal and a first order polynomial for the
background. Finally, the self-cross-feed contribution shows a peaking
component that is modeled as the product of a double Gaussian in \deltaE\
and a single Gaussian in \mes. The self cross-feed fraction
$f_{\rm scf} = \mbox{\SCFExpectedFractionNoErr}$, and the other
parameters that enter the definition of its contribution to the \pdf\ have
been determined from a Monte Carlo sample of signal events.

We vary the means of the narrow \deltaE\ and \mes signal
Gaussians, the coefficient in the exponential of the Argus
function, the linear coefficient of the \deltaE\ background distribution,
and the event yields $N_{\rm S}$ and $N_{\rm B}$. The means of the wide
Gaussians are determined by applying Monte Carlo determined offsets to the
mean of the narrow ones, such that only an overall shift of the fixed
\pdf\ shape is allowed.

Once the maximum likelihood fit provides the best estimates of the \pdf\
parameters, we use the \splot\ technique\ifthenelse{\boolean{physMode}}{
\cite{ref:Pivk_sPlot} }{
\cite{ref:Pivk_sPlot,ref:Pivk_sPlot_article,ref:Cahn_sPlot,ref:Snyder_sPlot}
} to reconstruct the efficiency-corrected \mlbp\ distribution and measure
the branching fraction.  The \pdf\ is used to compute the 
s-weight for the $n$-th component of event $e$ as:
 \begin{equation}
	\mbox{}_{s}{\cal{P}}_n(y_e) =
\frac{\sum_{j=1}^{n_{\rm c}}{{\bf V}_{nj} {\cal P}_{j}(y_e)}}
{\sum_{k=1}^{n_{\rm c}}{{N}_{k} {\cal P}_{k}(y_e)}}
\label{eqn:splot_weights}
\end{equation}
 where the indices $n,j$ and $k$ run over the $n_{\rm c}=2$ signal  and
background  components whose distributions, as functions of
$y_e=\left(\mes_{,e},\deltaE_e\right)$, are identified with the ${\cal
P}_{j}(y_e)$ symbol. ${\bf V}_{nj}$ is the covariance matrix of the event
yields as measured from the fit of the \pdf\ to the data sample. An
important property of the \splot~is that the sum of s-weights for the
signal or background component equals the corresponding number of fitted
signal or background candidates. Thus the \splot~is a good estimator of
the \mlbp\ distribution, and preserves the total signal yield, as
determined by the maximum likelihood fit. To retrieve the
efficiency-corrected number of events in given \mlbp\ bin $J$ we use the
s-weight sum:
 \begin{equation}
	N_{J} = \sum_{e\in
J}{\frac{\mbox{}_{s}{\cal{P}}_n(y_e)}{\varepsilon(x_e)}}
 \label{eqn:eff_corrected_n},
\end{equation}
 where $\varepsilon(x_e)$ is the per-event overall efficiency.
 The reconstruction efficiency depends on the position of the
candidate on the square Dalitz plane $x_e=\left(m_{\Lambda
p},\cos(\theta_{\rm H})\right)$, $\theta_{\rm H}$ being the helicity
angle of the pion in the $\bar{\Lambda} p$ rest frame, and has been
measured on a $10\times10$ grid
over the square Dalitz plane, using fully reconstructed 
signal Monte Carlo events.  The error 
$\sigma\left[N_{J}\right]$ in $N_{J}$ is given by:
 \begin{equation}
\sigma^2\left[N_{J}\right] =
\sum_{e\in J}{\left(\frac{\mbox{}_{s}{\cal{P}}_n(y_e)}{\varepsilon(x_e)}
\right)^2}.
\end{equation}
 An estimate of the efficiency-corrected number of events in the
sample is given by the sum of the efficiency-corrected s-weights or
 \begin{equation}
	N = \sum_{J}{N_{J}},
\end{equation}
 and the total branching fractions is obtained from 
\begin{equation}
{\cal{B}}\left(B\rightarrow \Lambda p \pi\right)
	= \frac{N}{N_{B\bar{B}}\cdot
{\cal{B}}\left(\Lambda\rightarrow p\pi\right)}.
\label{eqn:bratio_expr}
\end{equation}

Using fully reconstructed signal Monte Carlo events, we have checked that
this procedure provides a measurement of the \mlbp\ distribution and
total branching fraction with negligible biases and accurate errors.

\ifthenelse{\boolean{confFlag}}{
\section{SYSTEMATIC ERRORS}
}{
\section{Systematic errors}
}
\label{sec:Systematic_errors}

\begin{table}[t]
\caption[Breakdown of systematic errors]{Systematic 
errors on the BF measurement.}

\begin{center}
\begin{tabular}{|c c c|}
\hline
& source & error\\
\hline
Overall &  $B\bar{B}$ counting & \TotalBPairsSystematic \\
&$B^0\bar{B^0}/B\bar{B}$ fraction  & \BFractionSystematic \\
&Tracking efficiency & \TotalTrackSystematic \\
&PID efficiency & \PIDSystematic \\
&MC statistics & \DalitzStatSystematic \\
&$\Lambda\rightarrow p\pi$ branching fraction & \LambdaPPiBFSystematic \\
\hline
Event Selection&Event shape cut efficiency & \FDCutSystematicPerc\\
&Fit probability cut efficiency &  \VtxCutSystematicPerc\\
&$\Lambda$ flight length cut efficiency & \LambdaFlightSystematic\\
&$\Lambda$ mass cut efficiency & \LambdaMassSystematic\\
& $\Lambda_c$ veto cut & \LambdaCVetoSystematic\\
\hline
Fit Procedure&Likelihood parameters & \LikelihoodParameters\\
&\deltaE\ resolution& \DeltaESystematic\\
&Self cross-feed fraction & \SCFSystematic\\
&\splot\ bias correction& \LikelihoodBias\\

\hline
\hline
Total && \TotalSystematic\\
\hline
\end{tabular}
\label{tab:systematics}
\end{center}
\end{table}

Systematic errors are listed in Table I and classified as overall
uncertainties, uncertainties associated with event selection, and
uncertainties associated with fitting the signal event distribution.
Tracking efficiency uncertainty dominates the first category with a
contribution of \TotalTrackSystematic.
Particle identification systematic errors were evaluated by studying
data versus Monte Carlo agreement of identification efficiency on
protons from a pure sample of $\Lambda\rightarrow p \pi$ decays and pions
from $K^0_S\rightarrow \pi\pi$ decays. The finite signal Monte Carlo sample
available to measure the reconstruction efficiency over the
Dalitz plane, results in an additional \DalitzStatSystematic\ 
contribution to the systematic error. The uncertainty in
the determination of the number of $B\bar{B}$ pairs in the data sample
accounts
for a \TotalBPairsPerc\ systematic, while the assumption of a
\BZeroOverBChargedDefault\ ratio of $B^0\bar{B^0}$ to $B\bar{B}$
at the \FourS\  gives an additional \BFractionSystematic\ contribution,
computed as the difference with respect to the current measured value
\BZeroOverBChargedFraction\ \cite{ref:PDG}.

Event selection systematic errors associated with the determination of
the efficiencies of the Fisher-discriminant event shape cut and the vertex
fit-probability cut, have been evaluated comparing data
and Monte Carlo selection efficiencies of 
a sample of $B^0\rightarrow J/\psi K^0_S$ candidates.
In addition, we use an inclusive sample of $\Lambda\rightarrow p\pi$ 
candidates to estimate systematic errors associated with the
determination of efficiencies of flight length-significance cut
and $\Lambda$-mass requirement.

 The application of the requirement on the
reconstructed $m(\Lambda\pi)$ invariant mass to veto
$B^0\rightarrow\bar{\Lambda_c} p$ background has two associated systematic
effects. The first causes an approximate \LambdaCPiResBackSystematic\
increase in the branching fraction due to the residual $\Lambda_c$ component
that survives the cut. The second causes an approximate
\BLambdaPPiVetoSystematic\ reduction of the branching fraction due to the
reduced Dalitz-plot space. 
We take the larger of the two as the systematic error associated with the
$\Lambda_c$ veto cut.

We vary parameters
that are kept fixed in the likelihood fit by their statistical errors,
as measured on the signal MC sample fit, and measure the variation
of the \splot\ fitted result. The changes associated to the parameters that
enter the definition of the signal \pdf\ are conservatively considered
as fully correlated and added linearly to give a signal \pdf\ systematic
error of \SignalPDFSystematic, where the uncertainty on signal \mes\
fixed parameters
accounts for a \MesSystematic\ contribution and the uncertainty on
signal \deltaE\ fixed parameters for a \DESystematic\ contribution.
 The same procedure is applied to the
parameters that enter the background \pdf\ definition, with errors
determined on luminosity-weighted background MC samples, giving 
an additional \BackgroundPDFSystematic\ systematic error. Finally,
we combine in quadrature the two errors and obtain a 
\LikelihoodParameters\ systematic error
associated with uncertainties on the shape of signal and background 
\pdf\ models. 
The comparison of $B\rightarrow J/\psi K_S^0$ data and
Monte Carlo samples reveals that the width of the \deltaE\ Gaussian in the
signal \pdf\ can be underestimated in the Monte Carlo by up to
\DeltaEScaleDiffPerc, and this translates to an additional 
\DeltaESystematic\ systematic error associated with the
uncertainty in the \deltaE\ resolution.

We estimate possible biases associated with the determination of
yields with the \splot\ technique, using a collection of Monte Carlo
experiments in which signal candidates, generated and
reconstructed with a complete detector simulation, have been mixed with  
background candidates, choosing numbers of signal and background candidates
similar to those expected on the data sample under study.
Biases have been found within the statistical error in their measurement,
and we estimate a \LikelihoodBias\ systematic uncertainty associated
with the \splot\ fitting technique.

\ifthenelse{\boolean{confFlag}}{
\section{RESULTS}
}{
\section{Results}
}

\label{sec:Results}

\ifthenelse{\boolean{confFlag}}{
\begin{figure}[t]
\begin{center}
\includegraphics[width=7.0cm]
{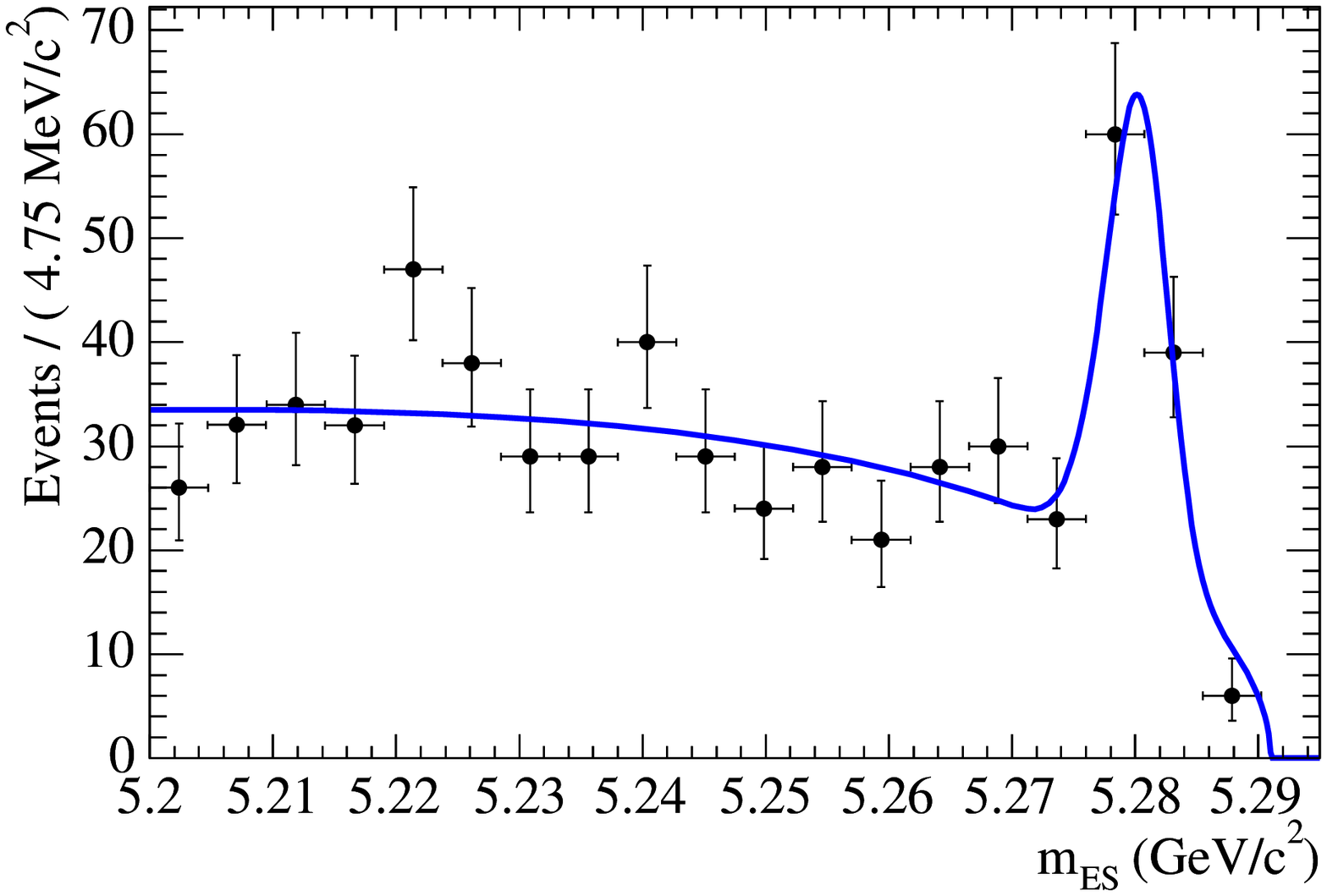}
\includegraphics[width=7.0cm]
{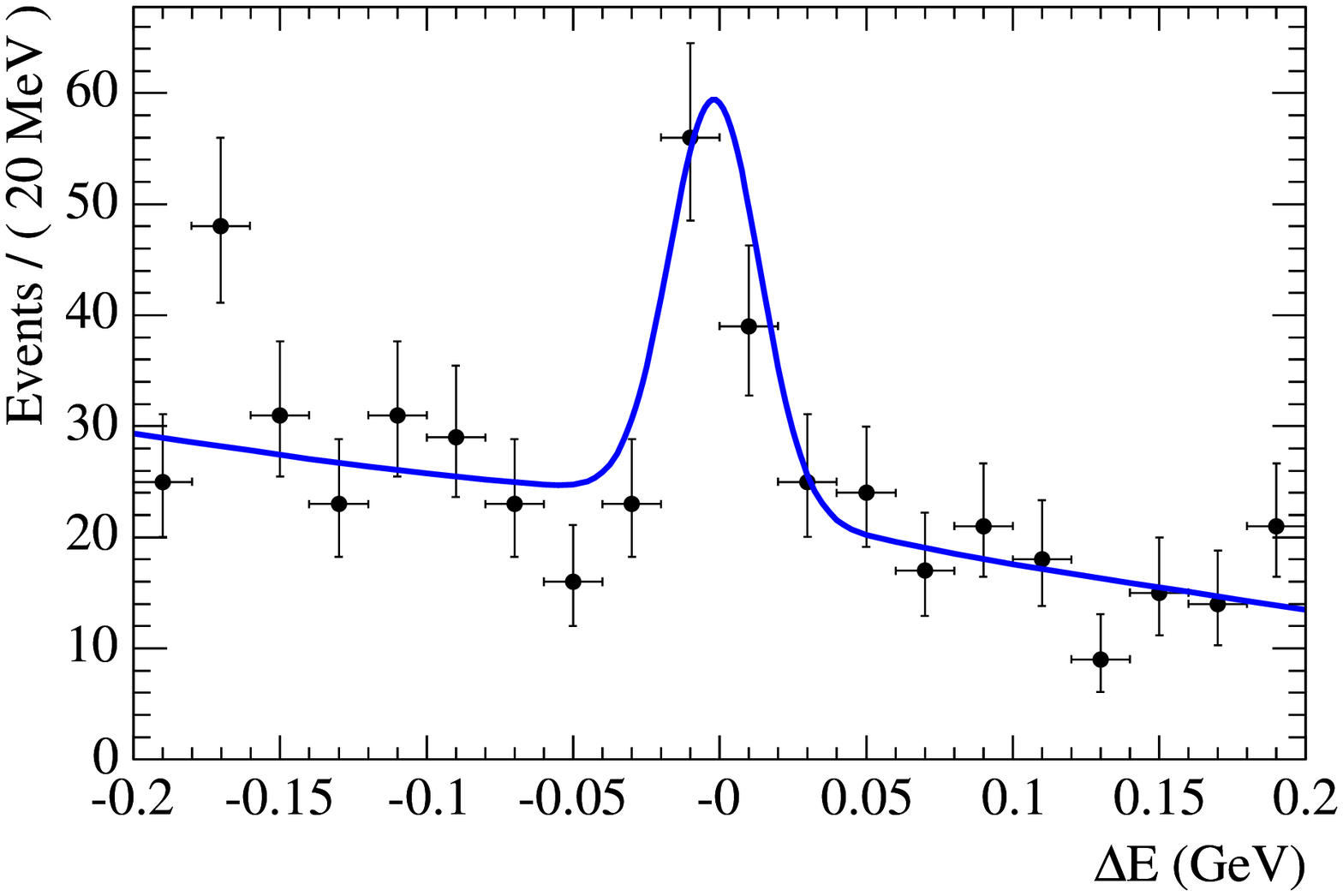}
\end{center}
\caption[Fit projections on \mes\ and \deltaE\ axes]{Left plot: 
\mes distribution of candidates with 
$|\deltaE|<\DeltaEWinCut\mev$. Right plot:
\deltaE\ distribution of candidates with
$\mes>\MesLowWinCut\gevcc$. Superimposed are projections
of the 2-dimensional fit \pdf\ onto the respective axes.}
\label{fig:Data_fit_Mes_DE_Projections}
\end{figure}
}{
\begin{figure}[t]
\begin{center}
\includegraphics[width=8.5cm]
{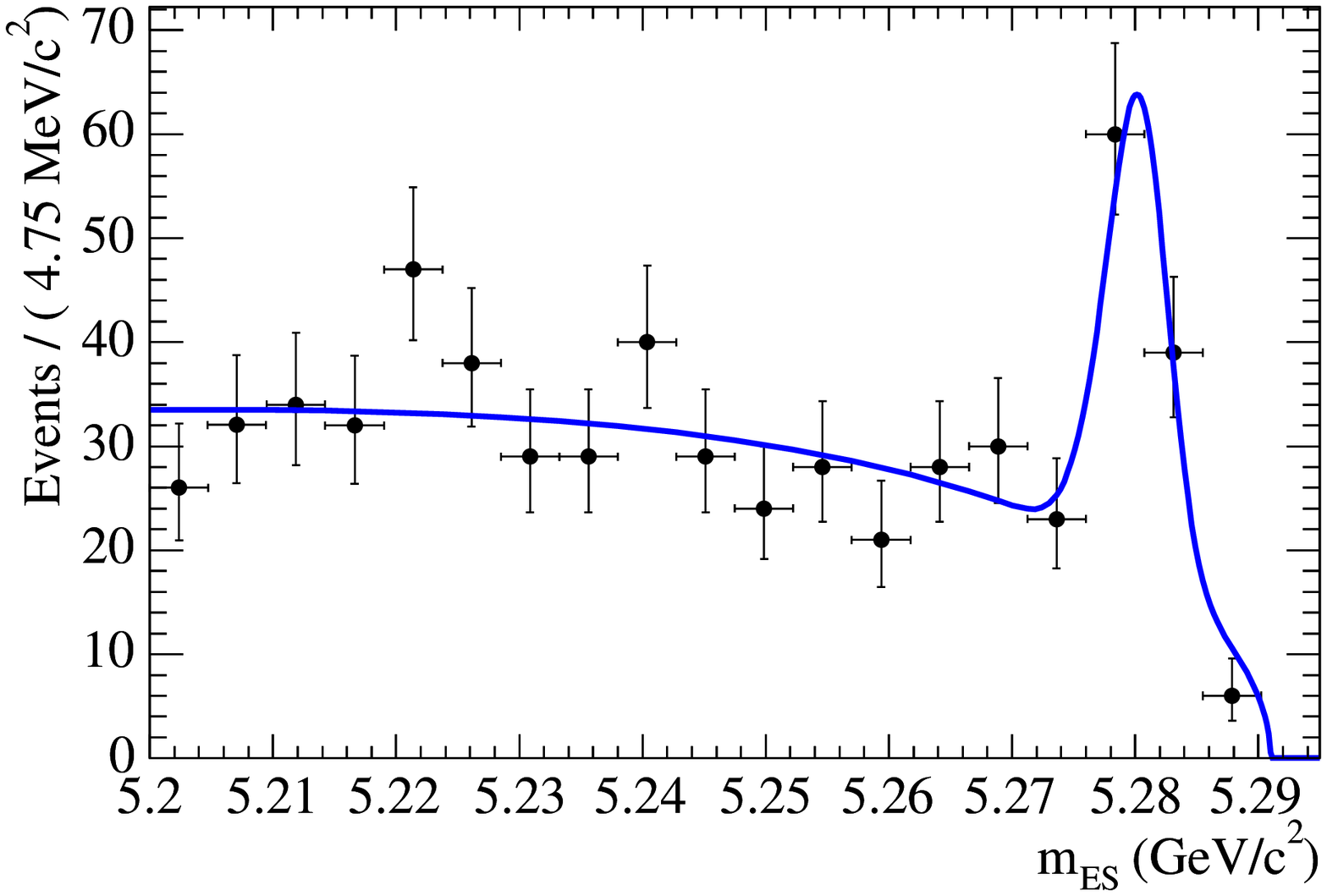}
\includegraphics[width=8.5cm]
{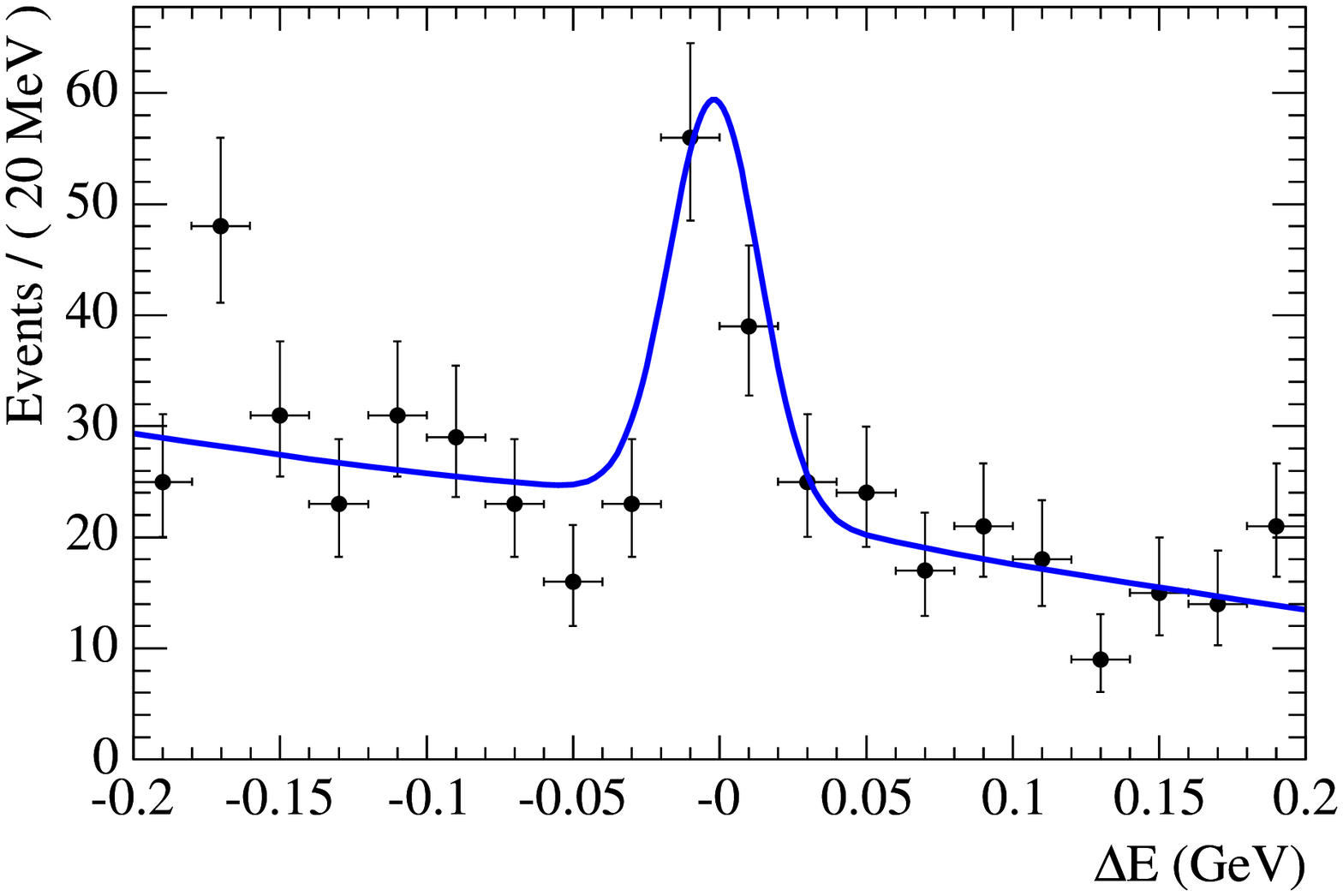}
\end{center}
\caption[Fit projections on \mes\ and \deltaE\ axes]{Upper plot: 
\mes distribution of candidates with 
$|\deltaE|<\DeltaEWinCut\mev$. Lower plot:
\deltaE\ distribution of candidates with
$\mes>\MesLowWinCut\gevcc$. Superimposed are projections
of the 2-dimensional fit \pdf\ onto the respective axes.}
\label{fig:Data_fit_Mes_DE_Projections}
\end{figure}
}

\begin{figure}[t]
\begin{center}
\includegraphics[width=8.5cm]
{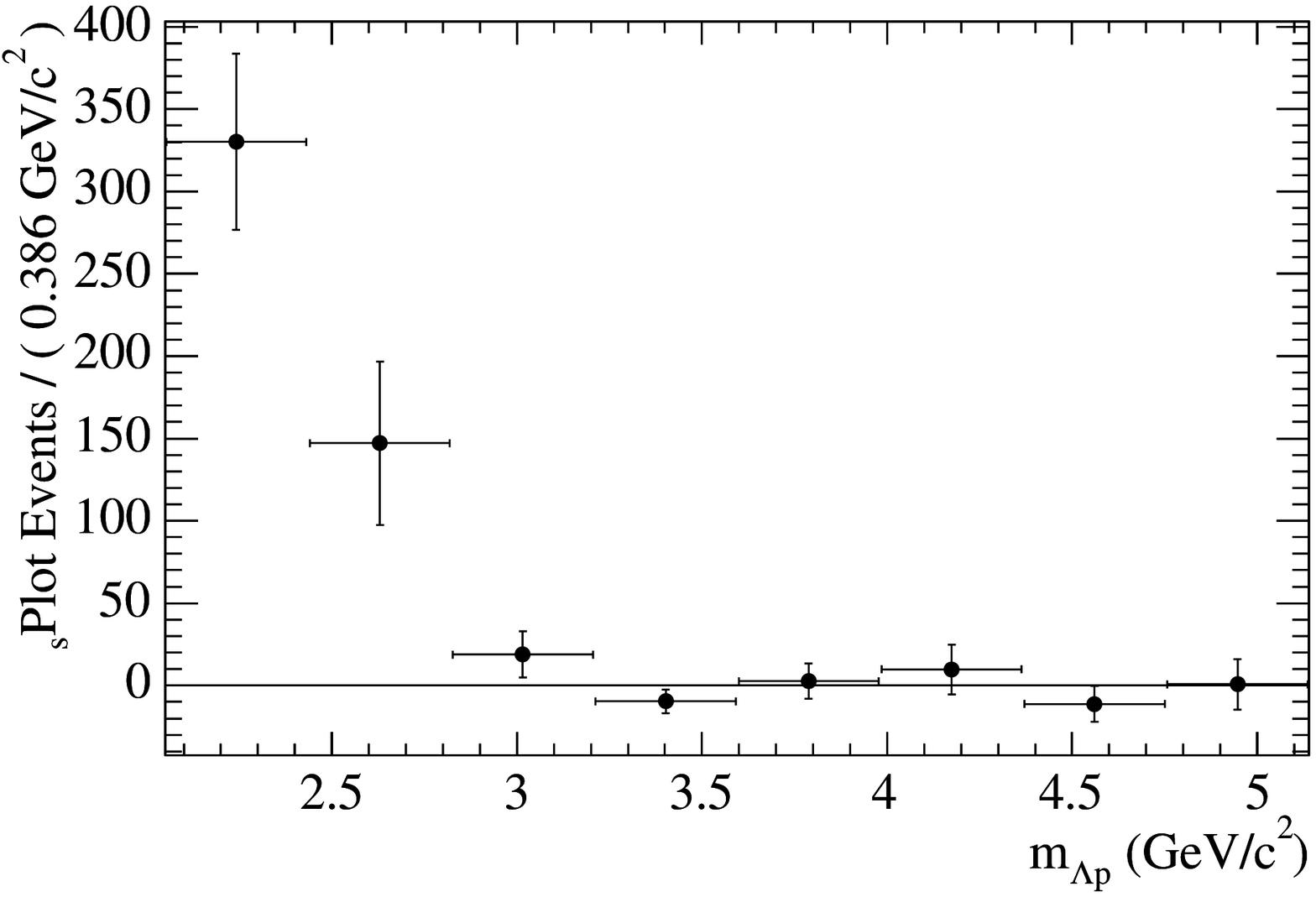}
\end{center}

\caption[\mlp\ event distributions]{\splot\ of the \mlp\ event
distribution with efficiency corrections
applied.}
\label{fig:Data_Mlp_Distributions}
\end{figure}

\begin{table}[t]
\caption[Likelihood fit results]{Likelihood fit result.
$N_S$ and 
$N_B$ are the number of fitted signal and background candidates respectively.
$\mu\left(\deltaE\right)$ is the mean value for the narrow Gaussian
of the $\deltaE$ signal \pdf\ component, while $c_1\left(\deltaE\right)$
is the slope of the linear $\deltaE$ background \pdf.
$\mu\left(\mes\right)$ is the mean value for the Gaussian of the 
\mes\ signal \pdf, and $c_{\rm Argus}\left(\mes\right)$ is the
coefficient at the exponent of the background \mes\ Argus function
as given in \cite{ref:Argus}. Reported errors are statistical only.
}
\begin{center}

\begin{tabular}{|c|c|}
\hline
Parameter & Value\\
\hline
$N_{\rm S}$ & \LikeNSFitResult\\
$N_{\rm B}$ & \LikeNBFitResult\\
\hline
$\mu\left(\deltaE\right)$ & \LikeMuDEFitResult \mev\\
$c_1\left(\deltaE\right)$ & \LikeCOneDEFitResult\\
\hline
$\mu\left(\mes\right)$ & \LikeMuMesFitResult\gevcc\\
$c_{\rm Argus}\left(\mes\right)$ & \LikeCArgusMesFitResult\\
\hline
\end{tabular}
\label{tab:LikelihoodFitResults}
\end{center}
\end{table}

We select a total of \DataSampleSelectedEvents\ candidates in the region
$|\deltaE|<\DeltaEFitRegion\mev$, $\mes>\MesFitRegion\gevcc$,
$|m(\Lambda\pi)-m(\Lambda_c)|>\LambdaCInvMassCut\mevcc$ in the
\TotalOnPeakLumi\invfb\ data sample considered. Table II contains the
fitted values of the 2-dimensional \mes-\deltaE\ \pdf\ parameters,
while Fig.~\ref{fig:Data_fit_Mes_DE_Projections} shows projections
of the 2-dimensional \pdf\ on the \mes\ and \deltaE\ axes.
Figure~\ref{fig:Data_Mlp_Distributions} shows the efficiency-corrected
signal \splot\ distribution of candidates as a function of the \mlbp\ coordinate;
this reveals a near-threshold enhancement similar to that observed in other
baryonic B decays.
Summing the efficiency-corrected \splot\ bins, we obtain a yield of 
$\mbox{\SPlotEffCorrectedYield}$ signal events, where the error is
statistical. Using
Equation~\ref{eqn:bratio_expr} we measure the branching fraction:
$${\cal{B}}(B^0\rightarrow \bar{\Lambda} p \pi^-) = \FinalBFCombined.$$
This measurement, which is compatible with a previous measurement by the
\belle\ collaboration\cite{ref:Belle_Blm_Update}, confirms the peaking
of the baryon-antibaryon mass spectrum toward threshold, a feature that
plays a key role in the explanation of the higher branching fraction
of three-body baryonic B decays with respect to two body ones.

\section{ACKNOWLEDGMENTS}

We are grateful for the 
extraordinary contributions of our \pep2\ colleagues in
achieving the excellent luminosity and machine conditions
that have made this work possible.
The success of this project also relies critically on the 
expertise and dedication of the computing organizations that 
support \babar.
The collaborating institutions wish to thank 
SLAC for its support and the kind hospitality extended to them. 
This work is supported by the
US Department of Energy
and National Science Foundation, the
Natural Sciences and Engineering Research Council (Canada),
Institute of High Energy Physics (China), the
Commissariat \`a l'Energie Atomique and
Institut National de Physique Nucl\'eaire et de Physique des Particules
(France), the
Bundesministerium f\"ur Bildung und Forschung and
Deutsche Forschungsgemeinschaft
(Germany), the
Istituto Nazionale di Fisica Nucleare (Italy),
the Foundation for Fundamental Research on Matter (The Netherlands),
the Research Council of Norway, the
Ministry of Science and Technology of the Russian Federation, 
Ministerio de Educaci\'on y Ciencia (Spain), and the
Particle Physics and Astronomy Research Council (United Kingdom). 
Individuals have received support from 
the Marie-Curie IEF program (European Union) and
the A. P. Sloan Foundation.

\ifthenelse{\boolean{authFlag}}{}{

\appendix
\newpage
\input{history_jour.tex}

}

\end{document}